# Erase-hidden and Drivability-improved Magnetic Non-Volatile Flip Flops with NAND-SPIN Devices

Ziyi Wang, Zhaohao Wang, *Member, IEEE*, Yansong Xu, Bi Wu, and Weisheng Zhao, *Fellow, IEEE*

*Abstract*—Non-volatile flip-flops (NVFFs) using power gating techniques promise to overcome the soaring leakage power consumption issue with the scaling of CMOS technology. Magnetic tunnel junction (MTJ) is a good candidate for constructing the NVFF thanks to its low power, high speed, good CMOS compatibility, etc. In this paper, we propose a novel magnetic NVFF based on an emerging memory device called NAND-SPIN. The data writing of NAND-SPIN is achieved by successively applying two unidirectional currents, which respectively generate the spin orbit torque (SOT) and spin transfer torque (STT) for erase and programming operations. This characteristic allows us to design an erase-hidden and drivability-improved magnetic NVFF. Furthermore, more design flexibility could be obtained since the backup operation of the proposed NVFF is not limited by the inherent slave latch. Simulation results show that our proposed NVFF achieves performance improvement in terms of power, delay and area, compared with conventional slave-latch-driven SOT-NVFF designs.

## I. INTRODUCTION

As key units in digital systems, flip-flops (FFs) store computing data for the central processing unit (CPU), micro controller unit (MCU), and so on. However, with CMOS technology scaling down, leakage power of the conventional flip-flops is becoming more and more significant [1][2]. In a microprocessor, the leakage power may exceed 40% of the total system power below 65 nm technology node [3]. Power gating (PG) technology, which makes the inactive circuits disconnected from the power supply in the standby mode, is a popular solution to eliminate the leakage power [4][5]. Nevertheless, the conventional CMOS FFs cannot retain the data during the power off, thus the data have to be backup into the main memory or disk drivers before triggering the PG, which results in additional speed and energy overhead. These bottlenecks can be overcome by designing a non-volatile flip-flop (NVFF) [6]-[11], which is a hybrid architecture combining the conventional CMOS FF with the emerging non-volatile memories (NVMs). While the PG is activated in a NVFF, the data can be backup into the local NVMs rather than be moved to the other modules. Thus, additional cost in the speed and energy could be avoided.

Amongst the NVFFs, magnetic NVFFs employing magnetic tunnel junction (MTJ) have been paid much attention due to high speed, low power, unlimited endurance and good CMOS compatibility [11]-[26]. Spin transfer torque (STT) is most widely used for the write operation of the MTJ [27]-[30], and hence for the backup operation of the magnetic NVFF. However, the performance of the STT-NVFFs is deteriorated by the asymmetry of backup operation between two write directions, which is attributed to the asymmetrical STT switching efficiency. In addition, the backup operation requires a pair of opposite currents to write '0' and '1', respectively, thus the source degeneration effect degrades the drivability of the related transistors, due to the different gate-source biases while writing '0' and '1' [31]. As a result, the transistors have to be large enough to tolerate the worst case of two switching events. However, it would induce overlarge current for the other case, which incurs larger energy dissipation and threatens the reliability of the devices. Although spin orbit torque (SOT) have been recently explored as an alternative to the STT [32]-[34], the source degeneration effect remains unresolved in the SOT-NVFF since the SOT switching current is still bi-directional.

In the most of the area-efficient magnetic NVFFs, the MTJs are directly switched by a pair of coupled CMOS inverters in the slave latch during the backup operation [21], [23], [25], [30]. Large-size transistors are required to ensure the drivability of the slave latch, which may degrade the performance of the CMOS part (i.e. volatile flip-flop structure). Moreover, most of the SOT-NVFFs employ two complementary SOT-MTJs to store 1-bit non-volatile data. The current channels (i.e. the heavy metal below the MTJ) have to be oppositely placed so that the MTJs are always written into the complementary states [11], [19], [23], [25], [26]. This type of layout will lead to difficulty in the practical fabrication process.

To overcome the aforementioned drawbacks of both STT-NVFFs and SOT-NVFFs, in this paper we propose an erase-

---

Ziyi Wang, Yansong Xu and Bi Wu are with the School of Electronics and Information Engineering, Beihang University, Beijing 100191, China.

Zhaohao Wang and Weisheng Zhao are with School of Microelectronics, Beijing Advanced Innovation Center for Big Data and Brain Computing (BDBC), Fert Beijing Research Institute, Beihang Univertisy, Beijing 100191, China (e-mail: zhaohao.wang@buaa.edu.cn; weisheng.zhao@buaa.edu.cn).

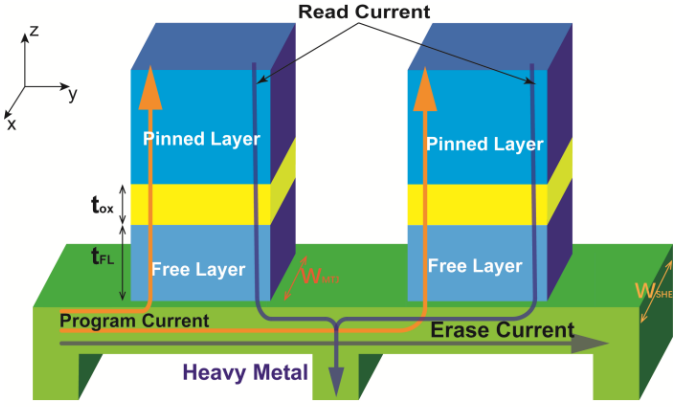

Fig. 1 Schematic structure and operation currents of a p-NANS-SPIN device.

hidden and drivability-improved magnetic NVFF based on a high-performance magnetic memory structure called NAND-SPIN [35]. Generally, the data is written into the NAND-SPIN by applying two unidirectional currents, which perform a SOT-induced erasing and a STT-driven programing, respectively. The erase operation is hidden into the active mode to reduce the backup delay. Furthermore, the source degeneration effect is eliminated thanks to the use of unidirectional switching currents. More importantly, the proposed NVFF relaxes the requirement on the drivability of the inherent slave latch, and therefore allows a more flexible design. Two embodiments of the proposed NVFF, which uses perpendicular- and in-plane-anisotropy MTJs, are designed and simulated with an electrical model of the SOT-MTJ and CMOS 40 nm design kit. Comparison among various NVFFs is shown based on the simulation results.

The remainder of this paper is organized as follows. Section II introduces the fundamentals of NAND-SPIN devices. In section III, we elaborate the circuit structure and operation of the proposed NVFF. The simulation results are discussed and evaluated in section IV. An extension work is described in section V. Finally, section VI concludes this paper.

## II. FUNDAMENTAL OF NAND-SPIN DEVICE

The NAND-SPIN device used in this work is composed of two MTJs and a heavy-metal strip, as shown in Fig. 1. An MTJ comprises of two ferromagnetic layers, called free layer (FL) and pinned layer (PL), separated by a tunneling oxide barrier. The magnetization of the PL is fixed whereas the FL magnetization can be switched between parallel and antiparallel (P and AP) to that of the PL, corresponding to low and high tunnel resistances, respectively. Two MTJs of the NAND-SPIN device are always switched to complementary resistance states, which represents 1-bit non-volatile data.

Here the MTJs are switched by a SOT-induced erasing and a STT-driven programing. The detailed operation is described as follows [35]. First, a current is applied to the heavy metal and induces the SOT, which switches both the MTJs to AP states. Thus, this step is called erase operation. Then the programming operation is performed. The SOT current is turned off, and another current flowing through one MTJ will switch it to P state by the STT. The other MTJ remains at AP state. As a result, two MTJs are switched to P and AP states, respectively. Depending on the relative states between two MTJs, the data '0' or '1' is stored.

Compared with the conventional STT- or SOT-MTJs, NAND-SPIN shows extraordinary properties in energy efficiency and area efficiency for three reasons. First, AP and P states are respectively written by the SOT and STT. Therefore, the write currents are unidirectional, which allows more compact circuit structure and avoids the source degeneration of the transistors [31]. Second, the asymmetry of the STT is alleviated since the STT is only used for the easier AP-to-P switching. Third, NAND-SPIN device could be fabricated as a whole by self-aligned process [36], which simplifies the procedure and improves the integration density.

An electrical model of NAND-SPIN is developed for the simulation and evaluation in this paper. The resistance of the MTJ is modeled by Brinkman model, bias-dependent tunneling magnetoresistance effect, Slonczewski model, etc. More details could be found in [37]. The magnetization switching is described by a Landau-Lifshitz-Gilbert (LLG) equation, which considers both the STT and SOT, as

$$\frac{\partial \boldsymbol{m}}{\partial t} = -\gamma \mu_0 \boldsymbol{m} \times \boldsymbol{H}_{\text{eff}} + \alpha \boldsymbol{m} \times \frac{\partial \boldsymbol{m}}{\partial t}$$
$$-\xi \wp J_{\text{STT}} \boldsymbol{m} \times (\boldsymbol{m} \times \boldsymbol{m}_p)$$
$$-\xi \eta J_{\text{SHE}} \boldsymbol{m} \times (\boldsymbol{m} \times \boldsymbol{\sigma}_{\text{SHE}}) \quad (1)$$

where $\boldsymbol{m}$ and $\boldsymbol{m}_p$ are the unit vectors along the magnetization orientation of the FL and PL, respectively. $\boldsymbol{\sigma}_{\text{SHE}}$ is the polarization direction of the SOT-induced spin accumulation. $J_{\text{STT}}$ and $J_{\text{SHE}}$ are the STT and SHE write current densities, respectively. $\boldsymbol{H}_{\text{eff}}$ is the effective magnetic field. $\gamma$ is the gyromagnetic ratio, $\mu_0$ is the vacuum permeability, $\xi = \gamma \hbar / 2 e t_F M_s$ with $\hbar$ the reduced Planck constant, $e$ the

TABLE I
SIMULATION PARAMETERS

| Parameters | iMTJ device [19][21][23][38] | pMTJ device [19][35][38][39] |
|---|---|---|
| Free layer volume | $\pi/4 \times 120\text{nm} \times 40\text{nm} \times 1.5\text{nm}$ | $40\text{nm} \times 40\text{nm} \times 0.7\text{nm}$ |
| Dimensions of W strip (w × l × d) | $180\text{nm} \times 60\text{nm} \times 5\text{nm}$ | $80\text{nm} \times 60\text{nm} \times 5\text{nm}$ |
| Damping constant ($\alpha$) | 0.0122 | 0.035 |
| Resistance-area product (RA) | $5\Omega \cdot \mu m^2$ | $5\Omega \cdot \mu m^2$ |
| Saturation magnetization ($M_s$) | 1000kA/m | 1000kA/m |
| Spin Hall angle ($\eta$) | 0.3 | 0.3 |
| Tunneling spin polarization (P) | 0.62 | 0.62 |
| TMR ratio | 120% | 120% |
| Heavy metal resistivity ($\rho_{HM}$) | $200\mu\Omega \cdot cm$ | $200\mu\Omega \cdot cm$ |
| Thermal stability barrier (E) | $92k_BT$ | $32k_BT$ |
| Asymmetrical STT efficiency ($\Lambda$) | 1.3 | 1.3 |
| Bias magnetic field ($B_{\text{ext}}$) | - | $-0.005/\mu_0$ |

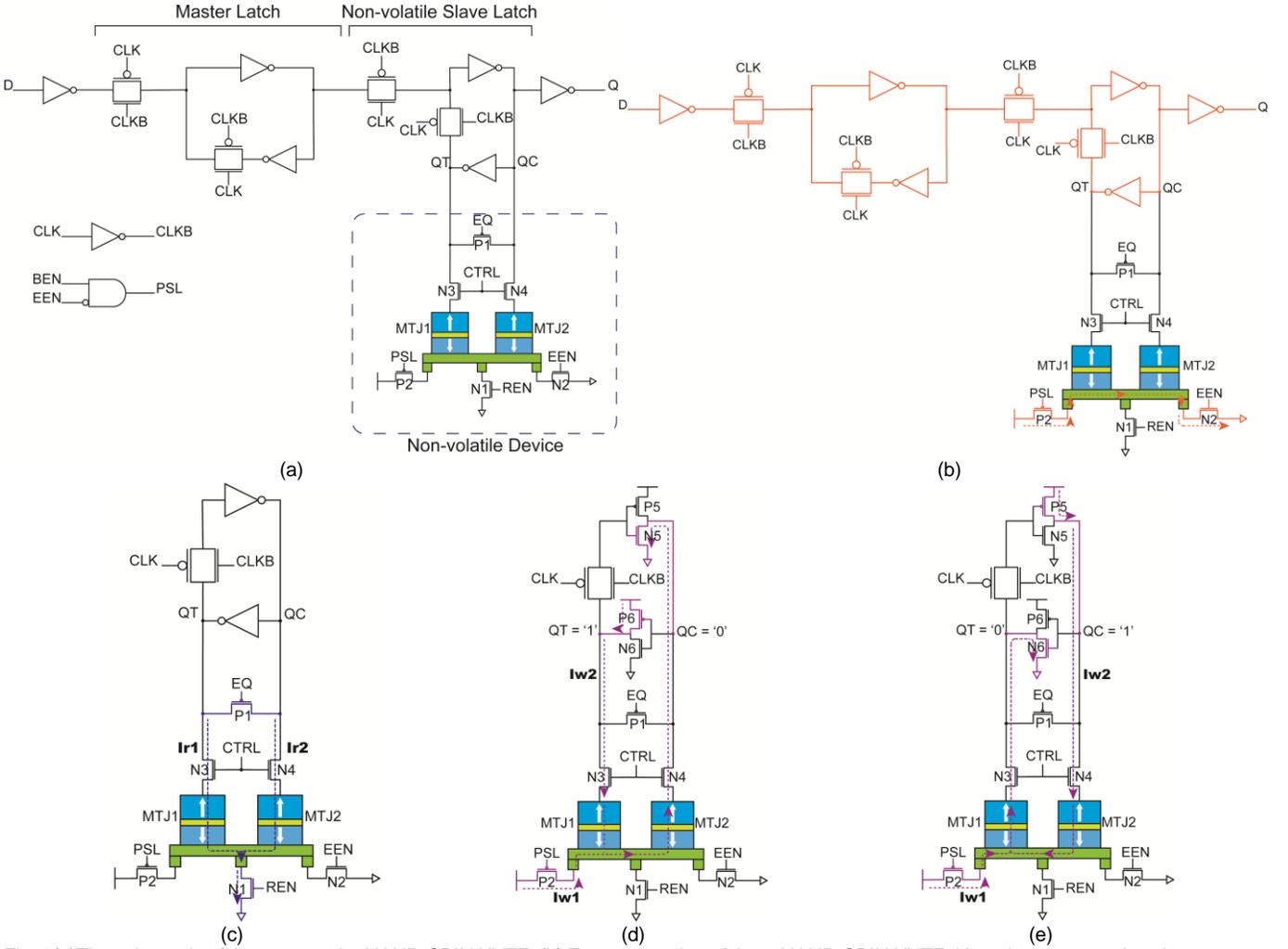

Fig. 2(a)The schematic of the proposed p-NAND-SPIN NVFF. (b) Erase operation of the p-NAND-SPIN NVFF. Here the orange-colored components are activated, i.e. the erase operation is hidden in the active mode. (c) Restore operation of the p-NAND-SPIN NVFF. (d) Backup '1' operation of the p-NAND-SPIN NVFF. (e) Backup '0' operation of the p-NAND-SPIN NVFF.

elementary charge, and $t_F$ the thickness of the FL. Other parameters are exhibited in Table I.

The asymmetry of the STT is reflected by $\wp$, as

$$\wp = \frac{2P\Lambda^2}{(\Lambda^2 + 1) + (\Lambda^2 - 1)(\boldsymbol{m} \cdot \boldsymbol{m}_p)} \quad (2)$$

where $\Lambda$ is a factor [38]. The larger $\Lambda$ induces more asymmetrical STT efficiency between two switching directions. $\Lambda = 1$ will eliminate the asymmetry of the STT.

## III. PROPOSED NVFF DESIGN WITH NAND-SPIN

This section presents the schematic and operation of the proposed NAND-SPIN based NVFF using both perpendicular- and in-plane-anisotropy MTJs (named as p-NAND-SPIN and i-NAND-SPIN). NVFFs targeting PG architectures could work in four modes: active, backup, standby, and restore modes. In the active mode, the NVFF operates as a normal volatile flip-flop (VFF). Before triggering the PG, the volatile data need to be written into the non-volatile NAND-SPIN in the backup mode. Then the power supply could be turned off in the standby mode to avoid the leakage power. Once the power supply is restarted, the non-volatile data inside the NAND-SPIN need to be written back to the volatile flip-flop in the restore mode. As mentioned in Section II, the data is written into the NAND-SPIN through an erase operation and a programming operation. It is worth noting that the polarity of the written data is determined by the programming operation rather than the erase operation. Therefore, we intentionally hide the erase operation inside the active mode so that only programming operation occurs in the backup mode. With this erase-hidden design, the backup latency is reduced, meanwhile the size of driving transistors could be tuned more flexibly.

### A. Circuit Operation

The schematic of the p-NAND-SPIN NVFF is illustrated in Fig. 2(a), including a CMOS-based VFF part (outside the dashed region) and a NAND-SPIN based non-volatile part (inside the dashed region). A slave latch is combined with one NAND-SPIN and six additional transistors (P1-2 and N1-4) for data retention during the standby mode. This non-volatile slave latch is connected with a clocked CMOS master latch to form the NVFF. The i-NAND-SPIN NVFF is not shown here for the simplicity, because it can be designed with the same structure as Fig. 2(a) except for the anisotropy direction of the MTJ.

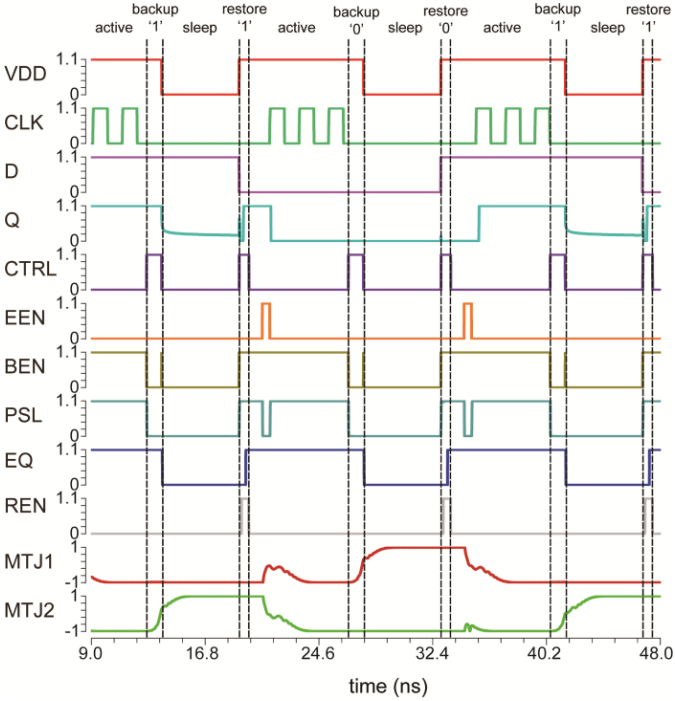

Fig. 3 Timing diagram of the proposed p-NAND-SPIN NVFF. The MTJ1 and MTJ2 signals are normalized z-component magnetization of the free layer. Others are voltage signals.

TABLE II
COMPARISON AMONG THE VARIOUS NVFFS.

| Parameter | | p-y type | i-y type | [23] | [25] | p-x type | i-x type |
|---|---|---|---|---|---|---|---|
| Delay (ns) | Backup | 1.04 | 1.87 | 4.95 | 4.32 | 0.99 | 1.868 |
| | Restore | 0.66 | 0.66 | 0.66 | 0.66 | 0.66 | 0.66 |
| Energy (fJ) | Backup | 84.3 | 216.5 | 281.7 | 239.5 | 84.1 | 216.4 |
| | Erase | 84.1 | 119.9 | - | - | 196.0 | 229.2 |
| | Restore | 56.0 | 54.0 | 67.6 | 53.0 | 58.4 | 64.4 |

The transistor sizes of PFET and NFET in VFF part of all NVFFs above are set to 400/40 and 290/40, respectively. In the proposed NVFFs (p-y type, i-y type, i-x type), the driving transistor P2 is set to 1000/40; in the proposed p-x type NVFF, the driving transistor P2 is set to 2000/40 to ensure a successful erase. The definitions of x-type and y-type will be described in Section V.

During the active mode, as shown in Fig. 2(b), EQ = '1', Ctrl = '0' and REN = '0', thus P1, N1 and N3-4 are deactivated. As a result, the NVFF acts as a CMOS-based VFF which is isolated from the NAND-SPIN. Once the NVFF just enters the active mode, transistors P2 and N2 is turned on for a short time to initialize the NAND-SPIN to state '11', which means that two MTJs are initialized to AP states (i.e. the erase operation mentioned above). This erase operation of the NAND-SPIN does not interfere with the normal operation of CMOS-based VFF due to the disconnected structure. As a result, the erase operation is hidden inside the active mode without additional latency.

Before the system enters the standby mode through the PG technology, the runtime data in the NVFF need to be backup into the NAND-SPIN.

To perform a backup, the CTRL signal and the PSL signal are asserted high, the CLK signal holds low for storing data into the slave latch. As mentioned above, both MTJ1 and MTJ2 have been erased to AP states during the active mode. Thus, one of two MTJs needs to be switched to P state for completing the backup. Assume QT = '1' and QC = '0', as shown in Fig. 2(d), transistors P2 and P6 work together as the current source, while transistor N5 acts as the current sink. Two currents (Iw1 and Iw2) driven by P2 and P6 are applied to the NAND-SPIN. As a result, MTJ2 is switched to P state through the STT, meanwhile MTJ1 remains at AP state. Similarly, if QT = '0' and QC = '1' (see Fig. 2(e)), MTJ1 and MTJ2 are respectively set to P and AP states. In summary, depending on the level of QT/QC, the two MTJs of the NAND-SPIN are programmed to complementary states 'P/AP' or 'AP/P' during the backup mode. It is important to mention that the STT is only responsible for AP-to-P switching, which requires smaller current and shorter latency compared to P-to-AP switching.

The restore operation is illustrated in Fig. 2(c). Once the power supply is restarted, EQ is kept at low level so that transistor P1 shorts QT and QC, and thus the power supply charge QT and QC to $V_{dd}/2$. Meanwhile the signal REN is pulled down and transistor N1 is open. Thus, QT and QC start discharging to form two read currents (Ir1 and Ir2) passing through MTJ1 and MTJ2, respectively. At this time the signal EQ is pulled up again to isolate QT from QC. As MTJ1 and MTJ2 have been written into complementary states during the backup mode, the discharging in the low MTJ resistance branch is faster than that in the other branch. As a result, the node on the low MTJ resistance branch drops to ground and the other node is pulled up to VDD. In this way, the complementary states in the NAND-SPIN are restored to the slave latch of VFF.

B. *Advantages of the Proposed NVFF*

Generally, most of the SOT-NVFF architectures rely on the inherent slave latch to achieve the backup operation. However, these architectures exhibit poor performance in backup delay and energy consumption. Below we will demonstrate how our proposed NVFF improves the speed and energy efficiency.

For the existing SOT-NVFFs, the polarity of the written data is dependent on the direction of the applied current. Therefore, bi-directional current need to be generated by the driving circuits. A common solution is to construct an independent write driver with additional four transistors [19], [20], but it leads to large area overhead. To achieve an area-efficient design, this bi-directional current is mostly provided by an inherent slave latch (e.g. transistors P5-6 and N5-6 in Fig. 2). To ensure sufficient write current, the transistors of the slave latch have to be large, which may degrade the performance of CMOS-based VFF part, such as large area overhead and strong parasitic effects. In contrast, our proposed NVFF employs unidirectional current to achieve the backup operation. Thus, the write current could be directly provided by the power supply rather than the slave latch. Only two access transistors (P2 and N2 in Fig. 2) are added for the efficient controlling. Furthermore, transistor P2 shows the largest drivability since it is directly connected to the power supply to maximize its source-gate bias (i.e. the source degeneration effect is eliminated). The same benefit is obtained by transistor N2 during the erase operation, as the unidirectional current is directly injected to ground. Thanks to this novel design, the current for the backup operation could be tuned by adjusting transistors P2 and N2, not limited by the size of the slave latch.

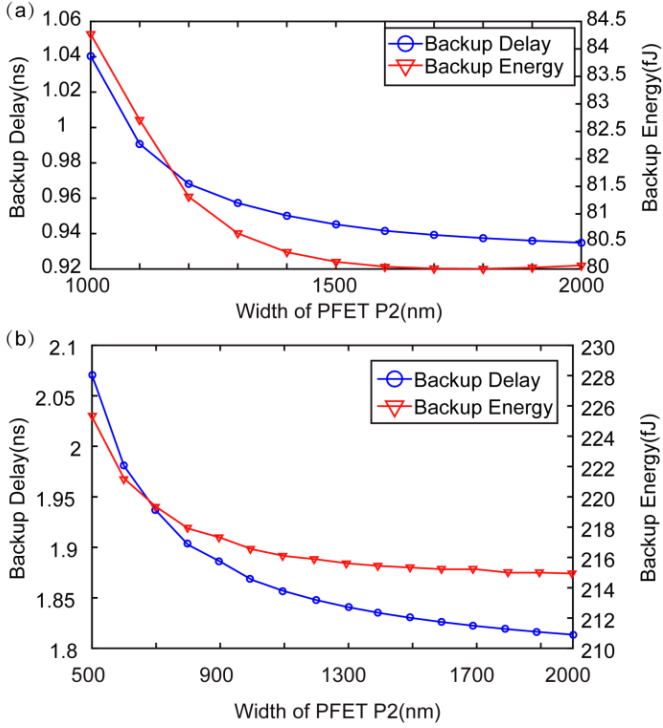

Fig. 4 Backup performance as a function of P2 transistor size with PFET/NFET of VFF set to 400/290. (a) p-NAND-SPIN NVFF (y-type) (b) i-NAND-SPIN NVFF (y-type).

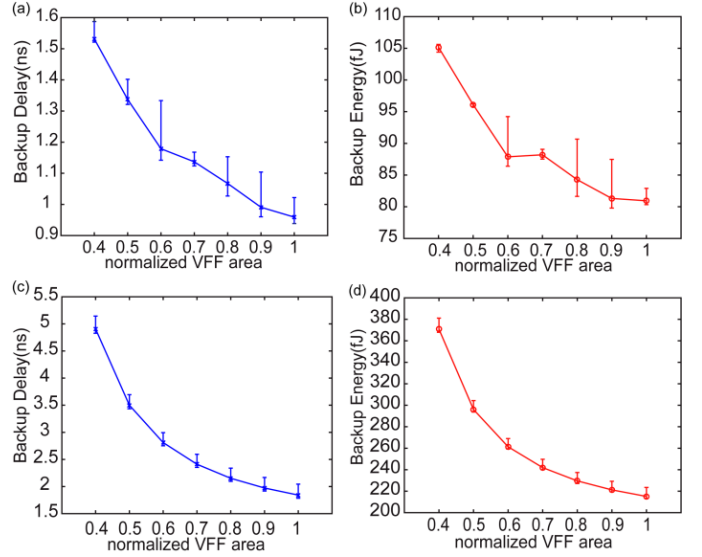

Fig. 5 The change of backup performance with the scaling of VFF area, and VFF area is normalized to the standard area with PFET/NFET = 400/300. Here the standard area is evaluated to be 1.6×3.9 μm$^2$ according to CMOS 40 nm design kit. (a)-(b) p-NAND-SPIN NVFF (y-type); (c)-(d) i-NAND-SPIN NVFF (y-type). The error bar indicates the range of backup performance while varying the P2 size.

In addition, some SOT-NVFFs employ a pair of oppositely-placed SOT-MTJs to store complementary states [11], [19], [23], [25], [26]. However, it causes the difficulty of manufacture since two SOT-MTJs have to be etched in the opposite position. In our proposed NAND-SPIN based NVFF, two MTJs share the same heavy metal strip which could be easily fabricated by self-aligned etching process [36].

## IV. RESULTS AND DISCUSSION

In this section the proposed NVFFs are simulated with a CMOS 40 nm design kit and a SOT-MTJ model mentioned in section II. The comparison of performance among various NVFFs is shown based on the simulation results. The optimization strategy of the proposed NVFF is discussed by investigating the roles of key parameters.

### A. Evaluation of Different SOT based NVFF Architectures

Fig. 3 shows the transient simulation results of the proposed p-NAND-SPIN NVFF. The signal waveforms reproduce the behaviors of the four operation modes described in section III-A. The timing diagram of i-NAND-SPIN NVFF can be obtained in a similar way, thus it would not be presented here for the simplicity.

Two NVFFs presented in [23] and [25] are chose as the baseline for comparison, because they showed more excellent performance than the existing other proposals. For a fair comparison, the same model and parameters are used among the various NVFFs. The detailed comparison results are listed in Table II.

The proposed i-NAND-SPIN NVFF design offers 10% lower backup energy with nearly 43% reduction of the backup delay compared to [25]. This improvement is mainly attributed to two reasons: first, in the i-NAND-SPIN NVFF the magnitude of the write current is not limited by the slave latch. Instead larger write current could be driven by the transistor P2. Second, only AP-to-P STT switching is performed in the backup mode, which costs smaller energy and shorter delay than the P-to-AP case due to the asymmetrical STT efficiency. For instance, P-to-AP STT switching delay of the proposed p-NAND-SPIN NVFF is as large as ~5.7 ns under the above parameter settings. Thus, in our proposed NAND-SPIN NVFF the AP state is written by the SOT-induced erase operation rather than the STT for avoiding the excessive cost in the speed and energy. In addition, the proposed i-NAND-SPIN offers 23% lower energy and 2.6× faster backup speed compared with [23], using the same operation mechanism.

As for the proposed p-NAND-SPIN NVFF, 65% lower backup energy and 4.2× faster backup speed can be achieved compared with [25]. The improvement is more significant than the case of i-NAND-SPIN NVFF, which is attributed to the fact that the perpendicular-anisotropy optimizes the STT efficiency.

It should be pointed out that concealment of erase operation also largely reduces delay and energy during the period of backup.

All the NVFFs mentioned above perform a restore operation in the same way, thus fewer differences can be seen from the comparison shown in Table II.

### B. Drivability analysis on the Proposed NVFF

As mentioned above, the currents for the backup operation mainly arise from the driving transistor P2 rather than the inherent slave latch. Therefore, more design flexibility could be

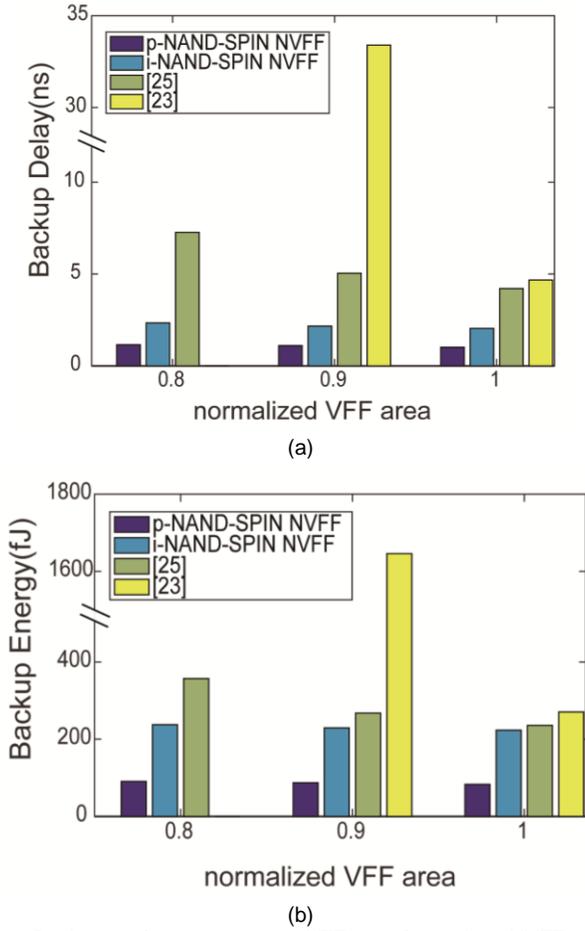

Fig. 6 Backup performance versus VFF area for various NVFFs. VFF area is normalized to the standard area with PFET/NFET = 400/300.

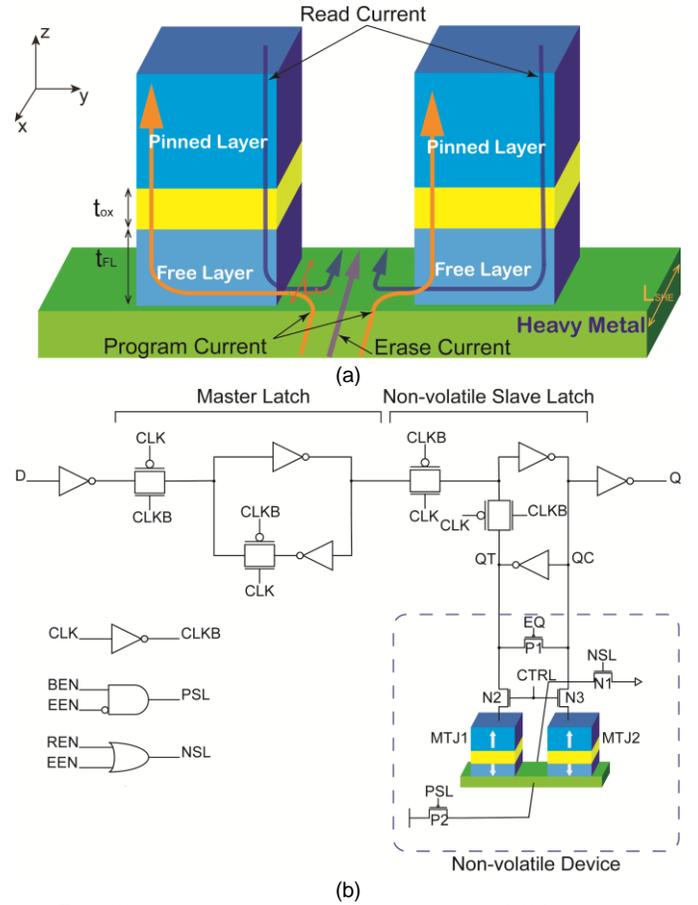

Fig. 7 (a) Schematic structure of a x-type p-NAND-SPIN device. (b) Schematic of a NVFF with x-type p-NAND-SPIN device.

obtained. Here we study the backup performance as a function of P2 transistor size, as shown in Fig. 4.

Given a fixed VFF area, we observe the change of backup performance by sweeping the width of the driving transistor P2 in both p- and i-NAND-SPIN NVFFs. As shown in Fig. 4(b), when the width of P2 varies from 500nm to 2000nm, the backup delay and energy of i-NAND-SPIN NVFF are reduced by 12.4% and 5%, respectively. Similarly, as shown in Fig. 4(a), they decrease by 10.2% and 5% when the width of P2 changes from 1000nm to 2000nm in the p-NAND-SPIN NVFF. This tendency could be easily understood since P2 acts as a current source offering the backup current. From the results, a trade-off could be obtained between the backup performance and P2 size: faster and more energy-efficient backup operation of NAND-SPIN NVFFs require a larger P2.

*C. Performance of the Proposed NVFF with CMOS-based VFF Scaling*

In addition to the driving transistor P2, the CMOS-based VFF part also has a non-negligible influence the backup performance. Figures 5-6 shows the backup performance as a function of the various VFF area. At each VFF area, a range of performance variation is obtained by adjusting the size of P2. It is seen that the backup performance is degraded with the scaling of the VFF, because the current source or current sink transistor is decreased to limit the total backup current. The degradation could be compensated by adjusting appropriately the P2 size, which is consistent with the conclusion of Fig. 4. For the p-NAND-SPIN NVFF, the ranges of variation at different VFF areas are not uniform, which could be attributed to the non-linear dependence of the switching speed on the current.

Although the backup performance is dependent on the VFF area, we argue that our NAND-SPIN NVFF has a more relaxed requirement on the VFF area. In contrast, previously proposed NVFFs are very sensitive to the VFF area, it can work only at a limited range of VFF area. In Fig. 6, we show the detailed comparison between our proposal and previous slave-latch-driven NVFFs. Clearly, the NVFF in [23] fails to perform a backup operation when the width of PFET/NFET in VFF is scaled down to 320/240 and 240/180, due to the inadequate driving current. The similar phenomenon could be observed from the NVFF in [25]. However, in our proposal, the unique characteristic of NAND-SPIN allows the use of P2 to provide additional drivability. Therefore, the backup operation is not easily blocked by the scaling VFF. In other words, our proposed NAND-SPIN NVFFs could achieve efficient backup operation in a wide range of VFF area. From instance, in Fig. 5 (c) and (d), when the PFET and NFET of VFF are respectively scaled from 400/40 and 300/40 to 240/40 and 180/40, the increment of backup delay and energy of i-NAND-SPIN NVFF are less than 1 ns and 50 fJ, respectively. These results mean that an area-efficient NVFF could be designed by scaling the VFF parts in our proposal.

## V. EXTENSION AND DISCUSSION

As an extension of the proposed NVFF, here we design a new type of NAND-SPIN device with another geometry, and evaluate its application in the NVFF.

### A. x-type NAND-SPIN

In Fig. 1, the erase current is applied to NAND-SPIN along y-axis. Actually, it is also feasible to apply it along x-axis, as shown in Fig. 7(a). Thus, we named them as y-type and x-type NAND-SPINs, respectively. It is worth noting that the exchange bias direction of x-type p-NAND-SPIN needs to be changed to x-axis for the deterministic switching. Similarly, the magnetization direction of x-type i-NAND-SPIN should be changed to y-axis.

The working mechanisms of x-type and y-type NAND-SPINs are identical, except for the direction of the applied current, as shown in Fig. 7(a). It is important to mention that the terminals of x-type NAND-SPIN is less than y-type NAND-SPIN, which means a smaller area overhead.

### B. x-type NAND-SPIN NVFF

The schematic of x-type NAND-SPIN NVFF are shown in Fig. 7(b). Compared to y-type NAND-SPIN NVFF, the programming currents for '0' and '1' are more symmetrical thanks to the altered geometry. The detailed performance metrics is listed in Table I. The erase operation of x-type NAND-SPIN NVFF requires longer delay, since the current is applied to a larger lateral area of the heavy metal for driving two MTJs in parallel. Nevertheless, the performance of NVFF is not deteriorated since the erase operation is hidden into the active mode. For the backup and restore performances, all the conclusions obtained from y-type NANS-SPIN NVFFs remain correct in the case of x-type NAND-SPIN NVFFs, due to the same operation mechanism.

## VI. CONCLUSION

We have proposed a novel NVFF by utilizing the advantages of the NAND-SPIN devices. The erase operation is hidden into the active mode for improving the operation speed. The unidirectional programming current of the NAND-SPIN devices allow to use an additional current source transistor to enhance the drivability, meanwhile relaxes the requirement on the VFF parts. Simulation results show the advantages of our proposed NVFF over the previous solutions. Moreover, more considerable tolerance to the VFF scaling is obtained in the proposed NAND-SPIN NVFF.

In this work we suppose that the programing operation of NAND-SPIN device is achieved by the sole STT mechanism. Actually, recent experiments have demonstrated that the programming current in the NAND-SPIN device inevitably passes the partial heavy metal and induces additional SOT effect, which is helpful to the magnetization switching [39], [40]. In addition, the current is non-uniformly distributed at the heavy-metal/MTJ interface, which assists the domain nucleation and reduce the switching current [40]. Therefore, all the above simulated results of the NAND-SPIN NVFF promise to be improved in the practical realization.

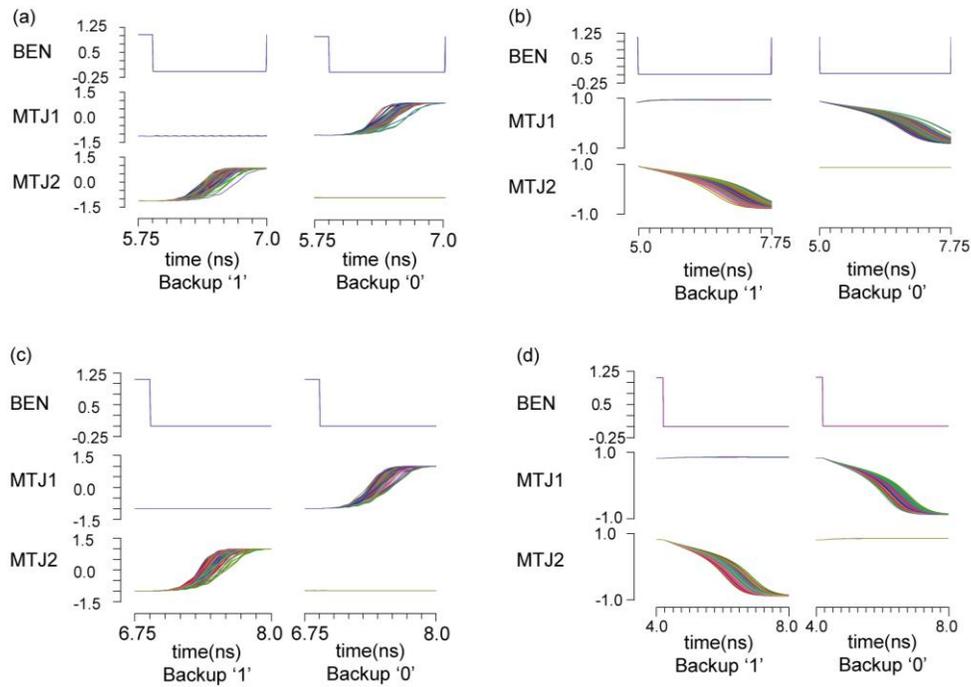

Fig. 8 Monte-Carlo simulation results of the backup operation. (a) p-y type (b) i-y type (c) p-x type (d) i-x type

TABLE III
MEAN AND STANDARD DEVIATION (STD DEV) OF QT, QC AND IW (IW1 + IW2) DURING THE BACKUP MODE.

|         | p-y backup 0 |        |        | p-y backup 1 |        |        |
|---------|--------|--------|--------|--------|--------|--------|
|         | QT (V) | QC (V) | Iw (A) | QT (V) | QC (V) | Iw (A) |
| Mean    | 124.1m | 1.1    | 78.9u  | 1.1    | 122.3m | 79.4u  |
| Std dev | 11.29m | 86.67u | 5.183u | 95.46u | 10.88m | 5.308u |
|         | i-y backup 0 |        |        | i-y backup 1 |        |        |
|         | QT (V) | QC (V) | Iw (A) | QT (V) | QC (V) | Iw (A) |
| Mean    | 180.7m | 1.1    | 104.5u | 1.1    | 178.6m | 104.2u |
| Std dev | 14.08m | 402.9u | 5.074u | 478.6u | 13.31m | 4.997u |
|         | p-x backup 0 |        |        | p-x backup 1 |        |        |
|         | QT (V) | QC (V) | Iw (A) | QT (V) | QC (V) | Iw (A) |
| Mean    | 125.1m | 1.1    | 79.79u | 1.1    | 125.5m | 80.13u |
| Std dev | 11.46m | 144.7u | 4.903u | 84.44u | 11.57m | 5.363u |
|         | i-x backup 0 |        |        | i-x backup 1 |        |        |
|         | QT (V) | QC (V) | Iw (A) | QT (V) | QC (V) | Iw (A) |
| Mean    | 175.2m | 1.1    | 102.8u | 1.1    | 175.5m | 103.1u |
| Std dev | 13.54m | 576.4u | 4.405u | 349.7u | 13.4m  | 4.905u |